\documentclass{article}

\usepackage[english]{babel}

\usepackage[letterpaper,top=2cm,bottom=2cm,left=3cm,right=3cm,marginparwidth=1.75cm]{geometry}

\usepackage{amsmath}
\usepackage{graphicx}
\usepackage{amssymb}
\usepackage[colorlinks=true, allcolors=blue]{hyperref}

\title{BinauralVAE: Spatial Audio Reconstruction For World Models}
\author{Luis Vitor Zerkowski \\
    VISGRAF \\
    IMPA \\
    \texttt{luisvz@gmail.com}
    \and
    Luiz Velho \\
    VISGRAF \\
    IMPA \\
    \texttt{lvelho@impa.br}
}

\begin{document}
\maketitle

\begin{abstract}
    Embodied artificial intelligence has historically very much relied on visual perception, leading to a proliferation of multiple vision-centric world models. However, this reliance fails to capture spatial understanding in its entirety and can even present vulnerabilities in environments with visual occlusions, low-light conditions, or blackouts-scenarios, where acoustic information becomes a critical alternative for spatial awareness and navigation. Despite its potential, research into realistic spatial audio and particularly the development of audio-centric world models remains sparse. In this technical report, we introduce BinauralVAE: a flexible, open-source pipeline (https://github.com/Luizerko/BinauralVAE) that explores multiple models for spatialized audio reconstruction, progressing from fundamental baselines to advanced, mathematically grounded architectures. Our approach evaluates various Variational Autoencoder architectures -- including complex-valued variants -- to learn robust latent representations of binaural signals. Developed alongside AudioWorldSim, our methodology leverages realistic acoustic data captured as a simulated robot navigates an environment. This pipeline establishes a foundation for state representation in a future audio-based world model, designed to map the direct causal connection between navigational actions and their resulting acoustic consequences, and helping to enable sound as an essential complementary modality for spatial knowledge acquisition.
\end{abstract}

\section{Introduction}

    The capacity to simulate and anticipate environmental dynamics is a central concept for autonomous agency. In recent years, this capability has been re-formalized through the development of world models, frameworks that compress high-dimensional observations into actionable latent spaces to facilitate reasoning, prediction, and planning \cite{ha2018world, li2025fromWwrds}. To date, however, the pursuit of artificial spatial intelligence to empower these world models has been dominated by visual modalities.
    
    The literature is rich with vision-centric examples: generalized architectures have expanded the application of world models across diverse and complex visual domains \cite{hafner2024masteringdiversedomainsworld}, while self-supervised learning on large-scale video data has enabled robust visual state prediction and planning \cite{assran2025vjepa2selfsupervisedvideo}. The field has even begun conceptualizing video generation models themselves as foundational world simulators \cite{openai2024videogeneration}. Moving beyond passive observation, recent generative techniques can now transform static images and videos into fully interactive, playable environments \cite{bruce2024geniegenerativeinteractiveenvironments, xia2024video2gamerealtimeinteractiverealistic, sam3dteam2026sam3d3dfyimages}. Additionally, within the realm of embodied AI, these visual models have been directly adapted for complex physical tasks, including robotic manipulation, navigational planning, and the stabilization of vision-language-action (VLA) policies \cite{huang2026pointworldscaling3dworld, bar2025navigationworldmodels, jiang2026wovrworldmodelsreliable}. Yet, while this extensive body of research represents monumental progress, it largely neglects sound, an essential dimension of spatial understanding.

    Biological perception is inherently multimodal, and constructing a comprehensive understanding of a physical space requires more than line-of-sight information. Sound naturally propagates around corners and obstacles, conveying information about scene geometry, material properties, and out-of-view dynamic events. Furthermore, relying exclusively on visual data introduces critical failure points in environments characterized by complex occlusions, low-light conditions, or total blackouts. Under these constraints, acoustic cues become even more of a vital mechanism for spatial awareness. By incorporating binaural spatial audio into world state representations, AI systems can progress to multisensory agents capable of true audio-visual imagination.

    Achieving this integration, however, is bottlenecked by scarcity of high-quality, dynamic binaural data tailored for active spatial understanding. While advanced simulators like ThreeDWorld \cite{gan2021threedworldplatforminteractivemultimodal} and SoundSpaces 2.0 \cite{chen2023soundspaces20simulationplatform} can render realistic multidirectional sound\footnote{For a comprehensive understanding of human auditory perception and computational audio processing, we recommend the audio overviews provided by Meta (https://developers.meta.com/horizon/design/audio/) and Unity (https://docs.unity3d.com/6000.5/Documentation/Manual/AudioOverview.html).}, extracting accessible, action-conditioned datasets at scale from these platforms has remained a significant challenge. To overcome this limitation, this project was developed alongside AudioWorldSim \cite{zerkowski2026audioworldsimrealisticbinauralaudio}, a dedicated data generation framework designed specifically to address this data scarcity, providing the exact state-action-audio dataset utilized in this work.

    Even when appropriate data is available, processing spatialized audio presents a challenge. Capturing realistic auditory dynamics requires fine-grained binaural information. Human spatial hearing relies on three primary mechanisms: Interaural Time Differences (ITD), which measure microsecond-level arrival delays between the ears; Interaural Level Differences (ILD), which capture intensity attenuation caused by the acoustic shadowing of the head; and spectral cues, which are individual frequency filterings shaped by a listener’s torso and outer ear for disambiguating sound sources along the vertical axis (up/down) and the front/back plane -- encapsulated computationally by a Head-Related Transfer Function (HRTF). Recent advances in deep learning have attempted to process all that using neural networks, predominantly by converting audio into magnitude spectrograms and processing them via Convolutional Neural Networks (CNNs) \cite{Purwins_2019, gong2021astaudiospectrogramtransformer, survey2023}. However, treating standard audio spectrograms merely as images is a fundamentally flawed assumption \cite{McLoughlin_2026}. More importantly, standard magnitude spectrograms discard phase information. Because spatial localization via ITD relies entirely on phase awareness, magnitude-only approaches are inherently inadequate for genuine spatial understanding.

    One method for overcoming this limitation is the use of complex-valued neural networks -- or complex-equivalent architectures -- that natively integrate phase information into their computations \cite{xie2025surveydeeplearningcomplex, lee2025binauralsoundeventlocalization, Tokala_2023, tokala2024binauralspeechenhancementusing, choi2019phaseawarespeechenhancementdeep}. In the generative domain, architectures such as Variational Autoencoders (VAEs) and their discrete latent variants (VQ-VAE) \cite{oord2018neuraldiscreterepresentationlearning} have proven highly effective at compressing and reconstructing rich audio representations. Alternatively, prominent audio encoders process raw waveforms directly, intrinsically preserving phase \cite{baevski2020wav2vec20frameworkselfsupervised, meta2020wav2vec2}. Yet, a critical misalignment persists because these phase-aware implementations are typically optimized for speech representation or source separation tasks. These acoustic distributions are fundamentally different from the reverberant, multidirectional environmental sounds an agent encounters during spatial navigation. Even within the domain of world models focused specifically on spatial navigation, recent works have begun to incorporate audio \cite{wang2026audiovisualworldmodelslearning, tian2026audioomniextendingmultimodalunderstanding, zeng2026semanticaudiovisualnavigationcontinuous}, but they often regress to spectrogram-only processing, inappropriately co-opt encoders trained on mismatched distributions, or remain closed-source regarding their exact auditory feature extraction.
    
    Consequently, open-source, phase-aware implementations designed specifically for spatial audio reconstruction in navigational world models remain practically non-existent. To bridge this gap, this technical report introduces BinauralVAE\footnote{https://github.com/Luizerko/BinauralVAE}, a flexible pipeline engineered to reconstruct spatialized audio captured during simulated robotic navigation. This work explores multiple VAE architectures, progressing from fundamental real-valued spectrogram baselines to a mathematically grounded, complex-valued VAE introduced by Nakashika et al. \cite{nakashika20_interspeech}. By mapping the direct causal connection between an agent's navigational actions and their resulting left- and right-ear acoustic consequences, we believe this framework establishes the encoding mechanisms required to elevate sound into a primary modality for spatial knowledge acquisition in world models.

\section{System Design}

    The foundation of this research relies on a flexible, end-to-end pipeline designed to process, train, and evaluate models for spatialized audio reconstruction. The system accommodates complete model training from scratch, automated hyperparameter grid searches, full-sequence audio inference, and latent space exploration -- what we refer to as "dreaming". Because of the architectural flexibility required to test multiple Variational Autoencoder (VAE) variants, the pipeline is heavily parameterized to ensure reproducible and modular experimentation.

    \subsection{Data Specification and Preprocessing}

        The pipeline ingests raw acoustic data generated by AudioWorldSim \cite{zerkowski2026audioworldsimrealisticbinauralaudio}. To map the direct causal connection between a robot's navigational actions and the resulting sound, the data is sliced into "binaural images." These are localized spectrograms representing exactly 0.2 seconds of audio, which precisely corresponds to the duration of a single simulated action. All audio is processed at a sample rate of 44.1 kHz with a hop length of 147.

        \begin{figure}[h]
            \centering
            \includegraphics[width=0.45\textwidth]{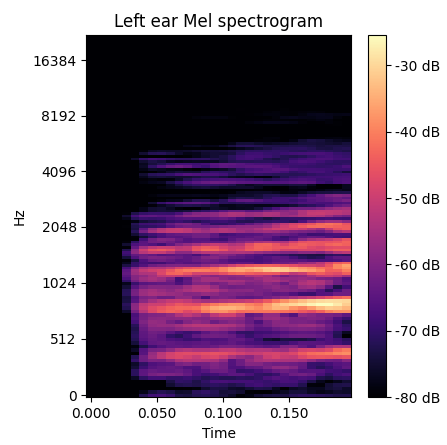}
            \hfill
            \includegraphics[width=0.45\textwidth]{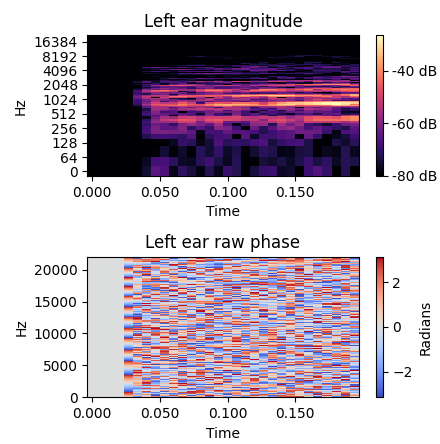}
            \caption{Comparison of "binaural images" split by discrete agent actions. On the left, a segment of Mel spectrogram. On the right, a segment of raw STFT output.}
            \label{fig:binaural_images}
        \end{figure}

        The preprocessing module then transforms and normalizes these data into one of three distinct modalities, tailored to the targeted neural architecture. The first modality, Mel STFT, computes a 2048-sample Short-Time Fourier Transform (STFT) mapped to 128 Mel bands, yielding a tensor of shape [2, 128, 60] that represents the left and right ear channels. This tensor is normalized using min-max scaling derived from global statistics across all navigation seeds. Additionally, the reference power is preserved to ensure better fidelity in reconstructions and dreaming downstream.
        
        The second modality, a 4-channel STFT, utilizes a 1024-sample STFT to produce a [4, 513, 60] tensor, independently stacking the magnitude and phase components for both ears. This representation is stabilized by applying a global min-max normalization and 0.3 power-law compression to the magnitude, alongside a $-\pi$ to $\pi$ normalization for the phase. Finally, the complex STFT modality retains the raw complex outputs from a 1024-sample STFT, generating a [2, 513, 60] tensor for the left and right channels. This representation is processed using maximum value normalization applied to the complex norm, combined with a 0.3 power-law compression on the magnitude -- a preprocessing step we found absolutely vital for stable network convergence.

    \subsection{Models}

        The primary objective of our pipelines is to reconstruct binaural audio segments corresponding to 0.2-second discrete robotic actions. By optimizing for high-fidelity reconstruction, we force the latent space to efficiently encode binaural acoustics, thereby forming the basis for latent state representations in a future audio-centric world model. To achieve this, we employ a Convolutional Neural Network (CNN) based VAE framework, directly inspired by the foundational visual architecture proposed by Ha and Schmidhuber \cite{ha2018worldmodels}. While simple, our CNN-based architecture is fast to implement and train, and crucially, makes it easier to trace successes and failures back to specific architectural changes. Furthermore, it can be flexibly adapted to process complex-valued acoustic signals. However, standard VAEs inherently optimize the maximum likelihood estimator of the data given the latent variables, effectively predicting the mean of a Gaussian distribution with an identity covariance. This formulation intrinsically penalizes variance, resulting in a well-documented smoothing effect that sacrifices fine, high-frequency textural details in the reconstructed outputs across all evaluated modalities.

        \subsubsection{Mel Spectrogram VAE}

            Our baseline approach processes Mel spectrograms using standard 2D convolutions, effectively treating the time-frequency representation as an image. A fundamental limitation of this methodology is the application of translation-equivariant convolution operations across the temporal axis. While this blurring of exact temporal events may be detrimental in tasks like speech recognition, for spatial audio understanding, the primary consequence is the destruction of the Interaural Time Difference (ITD).

            ITD refers to the microsecond-level delay between a sound wave reaching the left versus the right ear. In our dataset, audio is sampled at 44.1 kHz to satisfy the Nyquist criterion for the full range of human hearing. However, computing the Mel spectrogram with a 2048-sample window yields a temporal resolution of approximately 46 milliseconds per frame. This resolution is orders of magnitude too large to capture the microsecond-level phase delays required for ITD, rendering standard Mel spectrograms physically incapable of representing this vital temporal spatial cue.
            
            Despite this limitation, human spatial hearing also relies heavily on Interaural Level Differences (ILD) -- the intensity attenuation caused by acoustic head shadowing -- and spectral cues. These spectral cues, introduce specific frequency peaks and notches based on torso and pinna filtering, which are critical for disambiguating sound sources along the vertical (up/down) and median (front/back) planes. Because ILD and magnitude-based spectral cues are preserved within the Mel spectrogram, the VAE remains capable of learning a functional spatial representation.
            
\begin{figure}[h]
                \centering
                \includegraphics[width=0.8\textwidth]{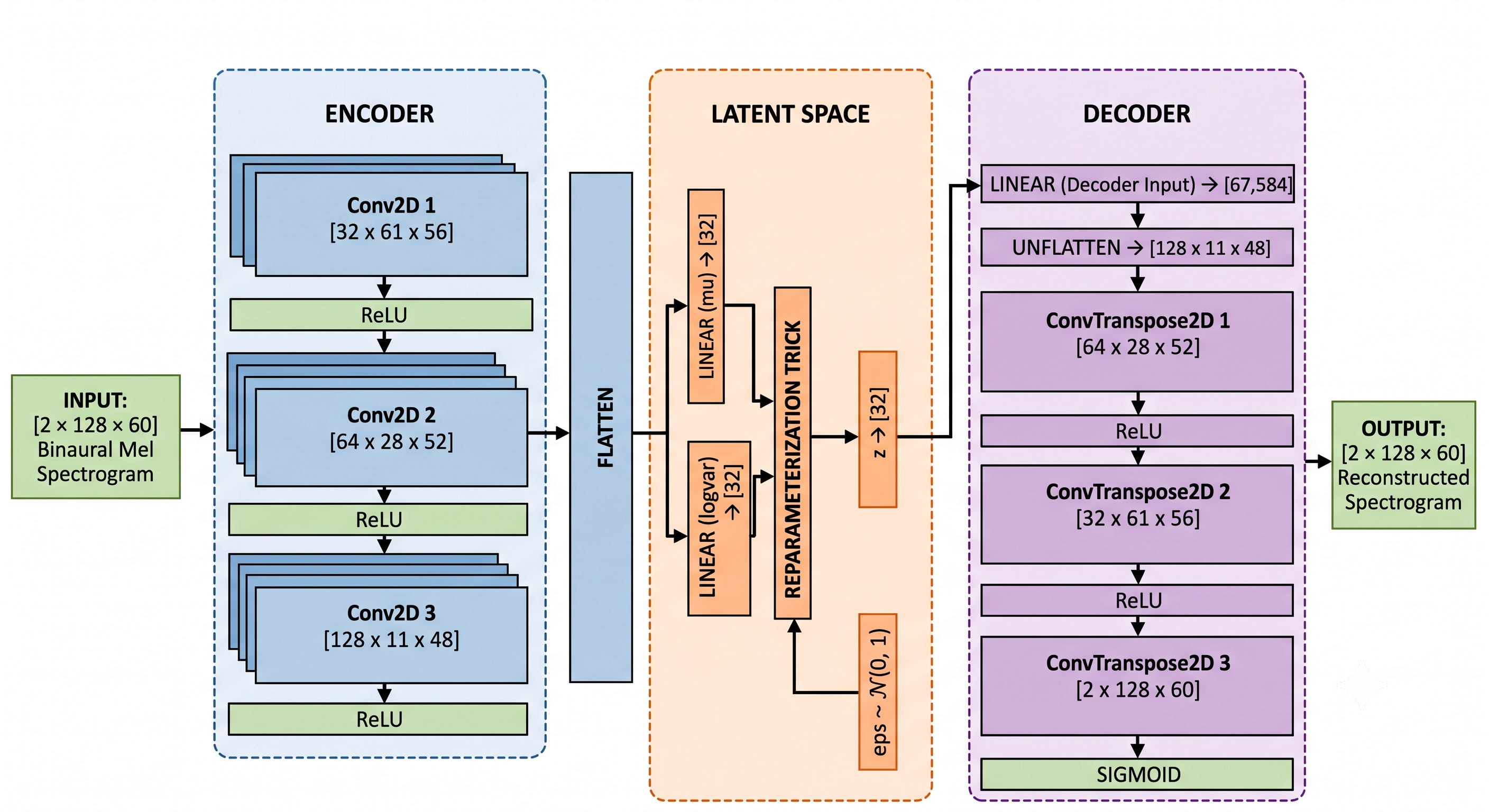}
                \caption{Architecture overview of the Mel Spectrogram VAE. The network accepts binaural Mel spectrograms as input, which are processed by an encoder consisting of three 2D convolutional layers paired with ReLU activations. The flattened feature map is passed through two parallel linear layers to predict the mean and log-variance of a 32-dimensional multivariate Gaussian posterior. Following the reparameterization trick, a sampled latent vector is passed to the decoder. The decoder reconstructs the signal using an initial linear layer followed by three 2D transposed convolutional layers with ReLU activations. The final transposed convolution, however, utilizes a Sigmoid activation instead of ReLU to bound the reconstructed output elements strictly within a $[0, 1]$ range.}
                \label{fig:mel_vae}
            \end{figure}

            The specific architecture employed for the Mel VAE is illustrated in Figure \ref{fig:mel_vae}. It is important to emphasize, however, that the underlying implementation is fully scalable and generalized, allowing researchers to easily customize the network depth and capacity for alternative configurations.
            
        \subsubsection{4-Channel STFT VAE}

            To address the loss of ITD inherent in the Mel spectrogram approach, our second architecture directly processes the magnitude and phase components of the raw STFT. By explicitly including phase information, the model gains access to the instantaneous frequency shifts between the left and right ears, theoretically restoring the necessary microsecond-level timing cues. Given this inclusion, we adjusted the STFT time-frequency trade-off by reducing the window size from 2048 to 1024 samples. While the magnitude-only Mel representation favored frequency resolution, this approach also relies on the phase to recover spatial cues, allowing the STFT to be biased toward a higher temporal resolution. 
            
            Despite the theoretical advantage of providing phase data, standard CNN-based architectures struggle significantly with phase reconstruction. This failure stems from two fundamental properties of acoustic phase. First, STFT phase resembles noise and lacks the smooth, predictable spatial gradients typical of image data. Accurately predicting it requires capturing highly variant, fine-grained details, a task at which standard VAEs fundamentally fail. Second, phase is a circular quantity that wraps from $-\pi$ to $\pi$. A standard neural network operating on linear feature representations fails to recognize the topological equivalence of $-\pi$ and $\pi$, resulting in massive penalty errors at the wrapping boundaries. Consequently, while the magnitude reconstruction can perform adequately, the phase reconstruction is bound to collapsing almost entirely. Unable to properly model the variance and heavily penalized by boundary errors, the network defaults to the safest optimization path to minimize the Mean Squared Error (MSE): it predicts a relatively flat, averaged value across the phase channels.
            
\begin{figure}[h]
                \centering
                \includegraphics[width=0.8\textwidth]{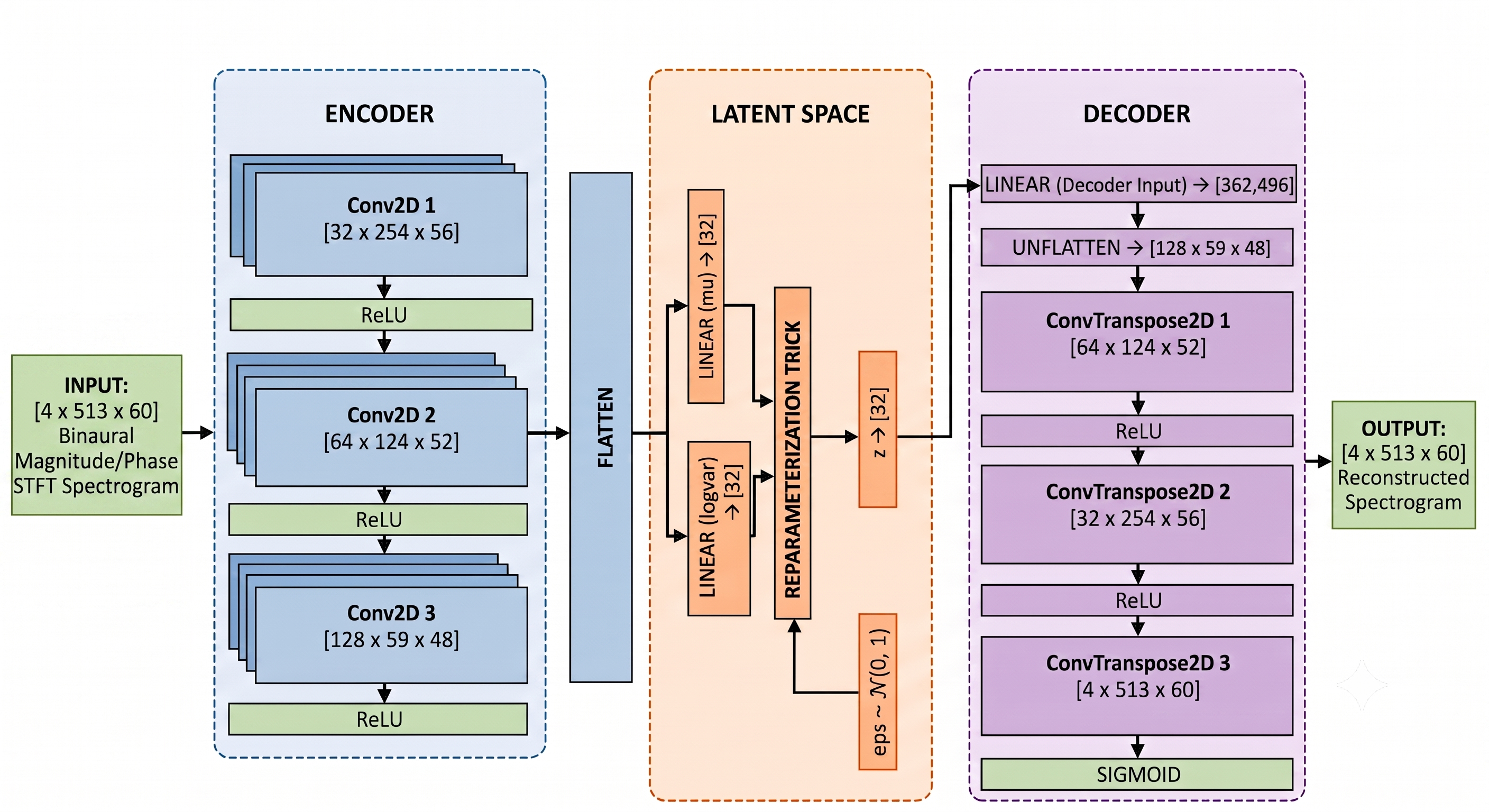}
                \caption{Architecture overview of the STFT 4-Channel Stacked VAE. The architecture accepts binaural STFT magnitude and phase as a 4-channel input. Aside from the differing input dimensions, the network topology is functionally identical to the Mel Spectrogram VAE.}
                \label{fig:stft_vae}
            \end{figure}

            The architecture for this approach is illustrated in Figure \ref{fig:stft_vae}. Aside from the dimensional adjustments required to accommodate the 4-channel input, the network topology is identical to the Mel VAE and remains fully scalable for further experimentation.
            
        \subsubsection{Complex VAE}

            Our third approach, the Complex-Valued VAE (CVAE), aims to achieve high-fidelity magnitude reconstruction while strictly preserving phase integrity. By transitioning from real-valued to complex-valued neural networks, the architecture intrinsically respects the complex nature of the STFT output. This paradigm naturally enforces the cyclic topology of acoustic phase, enabling improved spatial audio reconstructions. Our architecture builds upon the theoretical foundations established by Nakashika et al. \cite{nakashika20_interspeech}, who extended the standard VAE into the complex domain. This transition necessitates adapting all network parameters to maintain parallel real and imaginary weights, alongside expanding the underlying probabilistic framework to ensure that the prior, posterior, and likelihood functions are formulated as Complex Gaussian distributions.

            \vspace{1em}
            \noindent \textbf{Understanding Complex Gaussians}
            \vspace{0.5em}

                Unlike a standard real-valued Gaussian distribution, which is fully parameterized by a single covariance matrix, a Complex Gaussian distribution requires two distinct matrices to capture its full statistical profile: a covariance matrix $\mathbf{\Gamma} \in \mathbb{C}^{D \times D}$ and a pseudo-covariance matrix $\mathbf{C} \in \mathbb{C}^{D \times D}$. In a real-valued domain, a single matrix sufficiently describes both the variance of individual variables and their cross-correlations. However, because each variable in a Complex Gaussian occupies a two-dimensional subspace (real and imaginary), every pair of complex variables entails four distinct cross-correlations: real-to-real, real-to-imaginary, imaginary-to-real, and imaginary-to-imaginary. A single complex matrix provides only two degrees of freedom per variable pair, which is insufficient to map these four relationships. Therefore, both matrices are strictly required.

                Mathematically, the covariance matrix is defined using the conjugate transpose: $\mathbf{\Gamma} = \mathbb{E}[(\mathbf{z} - \mathbf{\mu})(\mathbf{z} - \mathbf{\mu})^H]$. Its diagonal elements are real, positive scalars representing the total variance ($\sigma_{real} + \sigma_{imag}$) -- effectively dictating the overall isotropic spread of the distribution. The off-diagonal elements are complex values that correlate distinct variables across both magnitude and phase. Conversely, the pseudo-covariance matrix is defined using the standard transpose: $\mathbf{C} = \mathbb{E}[(\mathbf{z} - \mathbf{\mu})(\mathbf{z} - \mathbf{\mu})^T]$. Here, the diagonal elements are complex values that capture the intrinsic correlation between the real and imaginary components of the same variable, allowing the distribution to form a skewed ellipse rather than a symmetric circle. Its off-diagonal elements capture the remaining cross-correlations between different complex variables.

                To formally illustrate this requirement, consider two zero-mean complex variables $z_1 = x_1 + iy_1$ and $z_2 = x_2 + iy_2$. The covariance between these variables (the off-diagonal elements) is formulated as $\mathbb{E}[z_1 z_2^{\ast}] = \mathbb{E}[(x_1 + iy_1)(x_2 - iy_2)] = \mathbb{E}[x_1 x_2 + y_1 y_2 + i(y_1 x_2 - x_1 y_2)]$. By defining the four core real-valued cross-correlations as $A = \mathbb{E}[x_1 x_2]$, $B = \mathbb{E}[y_1 y_2]$, $D = \mathbb{E}[y_1 x_2]$, and $E = \mathbb{E}[x_1 y_2]$, the complex covariance can be expressed as $\mathbb{E}[z_1 z_2^{\ast}] = (A + B) + i(D - E)$. Similarly, the pseudo-covariance is formulated as $\mathbb{E}[z_1 z_2] = \mathbb{E}[(x_1 + iy_1)(x_2 + iy_2)] = \mathbb{E}[x_1 x_2 - y_1 y_2 + i(y_1 x_2 + x_1 y_2)]$, which simplifies to $\mathbb{E}[z_1 z_2] = (A - B) + i(D + E)$. Because there are four unknown real cross-correlations ($A, B, D, E$), both the covariance and pseudo-covariance are mathematically necessary to solve the system. Specifically, $(A + B)$ and $(A - B)$ isolate $A$ and $B$, while $(D - E)$ and $(D + E)$ isolate $D$ and $E$.

                Applying this logic to a single zero-mean complex variable $z = x + iy$, we can derive the diagonal elements of these matrices. The total variance is given by $\sigma = \mathbb{E}[zz^{\ast}] = \mathbb{E}[(x+iy)(x-iy)] = \mathbb{E}[x^2 + y^2] = \sigma_{xx} + \sigma_{yy}$. This results in a real, positive scalar representing the total isotropic spread. The pseudo-variance is given by $\delta = \mathbb{E}[zz] = \mathbb{E}[(x+iy)(x+iy)] = \mathbb{E}[x^2 - y^2 + 2ixy] = (\sigma_{xx} - \sigma_{yy}) + 2i\sigma_{xy}$. This yields a complex number ($\delta_r + i\delta_i$) that captures both the variance imbalance ($\sigma_{xx} - \sigma_{yy}$) and the covariance ($\sigma_{xy}$) between the real and imaginary components, defining the elliptical skew of the distribution. By combining $\sigma$ and $\delta$, the underlying $2 \times 2$ real covariance matrix $\mathbf{S}$ for the $x$ and $y$ components of a single complex variable is perfectly reconstructed:
                
                $$
                \mathbf{S} = 
                \begin{bmatrix}
                    \sigma_{xx} & \sigma_{xy} \\
                    \sigma_{yx} & \sigma_{yy}
                \end{bmatrix} = 
                \begin{bmatrix}
                    \frac{\sigma + \delta_r}{2} & \frac{\delta_i}{2} \\
                    \frac{\delta_i}{2} & \frac{\sigma - \delta_r}{2}
                \end{bmatrix}
                $$

            \vspace{1em}
            \noindent \textbf{CVAE Formulation and KL-Divergence}
            \vspace{0.5em}

                Building upon the definitions of complex Gaussian distributions, we formulate the probabilistic framework of the CVAE. We begin by defining the prior distribution. Analogous to the standard VAE, which utilizes an isotropic real-valued Gaussian prior $p(\mathbf{h}) = \mathcal{N}(\mathbf{0}, \mathbf{I})$, the CVAE employs a standard complex normal prior $p(\mathbf{h}) = \mathcal{N}_c(\mathbf{0}, \mathbf{I}, \mathbf{O})$. The zero pseudo-covariance matrix ($\mathbf{O}$) indicates that the complex latent variables are mutually uncorrelated and individually form perfectly symmetric, isotropic unit circles in the complex plane.

                For the approximate posterior, a standard VAE assumes a factorized Gaussian $q_{\phi}(\mathbf{h}\vert{}\mathbf{z}) = \mathcal{N}(\mathbf{\mu}, \Delta(\mathbf{\sigma}))$, where $\Delta(\cdot)$ denotes a diagonal matrix. This structural assumption ensures mathematical tractability and enforces disentanglement, forcing each latent dimension to encode independent features. Applying this exact inductive bias to the complex domain, we define the complex posterior as $q_{\phi}(\mathbf{h}\vert{}\mathbf{z}) = \mathcal{N}_c(\mathbf{\mu}, \Delta(\mathbf{\sigma}), \Delta(\mathbf{\delta}))$. By restricting both the covariance and pseudo-covariance to diagonal matrices, we enforce strict independence across different complex variables. However, the inclusion of the pseudo-variance term $\delta_j$ allows each individual latent variable to form a skewed, elliptical distribution across its own real and imaginary components.

                Given this formulation, we can derive the Kullback-Leibler (KL) divergence between the approximate posterior and the prior. Because the latent variables are strictly independent, the total KL-divergence is simply the sum of the divergences across each individual latent dimension $h_j$. We compute the divergence for each $h_j$ by treating its corresponding prior and posterior as bivariate (2D) real-valued Gaussians. The general KL-divergence formula between a $k$-dimensional posterior $q = \mathcal{N}(\mathbf{\mu}_q, \mathbf{\Sigma}_q)$ and prior $p = \mathcal{N}(\mathbf{\mu}_p, \mathbf{\Sigma}_p)$ is defined as:
                
                $$D_{KL}(q \vert{}\vert{} p) = \frac{1}{2} \left[ \text{tr}(\mathbf{\Sigma}_{p}^{-1} \mathbf{\Sigma}_{q}) + (\mathbf{\mu}_q - \mathbf{\mu}_p)^T \mathbf{\Sigma}_{p}^{-1} (\mathbf{\mu}_q - \mathbf{\mu}_p) - k + \ln\left(\frac{\vert{}\mathbf{\Sigma}_{p}\vert{}}{\vert{}\mathbf{\Sigma}_{q}\vert{}}\right) \right]$$

                To evaluate this for a single complex variable $h_j$ (where $k=2$, representing the real and imaginary components), we establish the parameters for both distributions:

                \begin{itemize}
                    \item The Prior ($p_j$): As a standard complex normal, its total unit variance is distributed equally across the real and imaginary axes, yielding $\mathbf{\mu}_p = \mathbf{0}$ and $\mathbf{\Sigma}_p = \frac{1}{2}\mathbf{I}$. Consequently, its inverse is $\mathbf{\Sigma}_p^{-1} = 2\mathbf{I}$ and its determinant is $\vert{}\mathbf{\Sigma}_p\vert{} = \frac{1}{4}$.

                    \item The Posterior ($q_j$): The posterior has a mean vector $\mathbf{\mu}_q = [\text{Re}(\mu_j), \text{Im}(\mu_j)]^T$. As derived previously, its $2 \times 2$ real covariance matrix is:

                    $$
                    \mathbf{\Sigma}_q = \mathbf{S}_j = 
                    \begin{bmatrix} 
                        \frac{\sigma_j + \text{Re}(\delta_j)}{2} & \frac{\text{Im}(\delta_j)}{2} \\ 
                        \frac{\text{Im}(\delta_j)}{2} & \frac{\sigma_j - \text{Re}(\delta_j)}{2}
                    \end{bmatrix}
                    $$
                    
                \end{itemize}
                
                The determinant of this covariance matrix simplifies to $\vert{}\mathbf{\Sigma}_q\vert{} = \frac{\sigma_j^2 - \vert{}\delta_j\vert{}^2}{4}$.
                
                By substituting these parameters into the KL-divergence equation, we evaluate each term inside the brackets:

                \begin{itemize}
                    \item $\text{tr}(2\mathbf{I} \cdot \mathbf{\Sigma}_q) = 2 \left( \frac{\sigma_j + \text{Re}(\delta_j)}{2} + \frac{\sigma_j - \text{Re}(\delta_j)}{2} \right) = 2\sigma_j$

                    \item $\mathbf{\mu}_q^T (2\mathbf{I}) \mathbf{\mu}_q = 2 (\text{Re}(\mu_j)^2 + \text{Im}(\mu_j)^2) = 2\vert{}\mu_j\vert{}^2$

                    \item $-k = -2$

                    \item $\ln\left(\frac{1/4}{(\sigma_j^2 - \vert{}\delta_j\vert{}^2)/4}\right) = \ln\left(\frac{1}{\sigma_j^2 - \vert{}\delta_j\vert{}^2}\right) = -\ln(\sigma_j^2 - \vert{}\delta_j\vert{}^2)$
                \end{itemize}
                
                Combining these terms and multiplying by the leading $\frac{1}{2}$ produces the KL-divergence for a single complex variable $h_j$:
                
                $$
                D_{KL}(q_j \vert{}\vert{} p_j) = \frac{1}{2} \left[ 2\sigma_j + 2\vert{}\mu_j\vert{}^2 - 2 - \ln(\sigma_j^2 - \vert{}\delta_j\vert{}^2) \right] = 
                $$
                
                $$ = \vert{}\mu_j\vert{}^2 + \sigma_j - 1 - \frac{1}{2}\log(\sigma_j^2 - \vert{}\delta_j\vert{}^2)
                $$
                
                Finally, the total KL-divergence for the entire latent space is obtained by summing over all latent dimensions. Expressed in vector notation, this simplifies to:

                $$
                D_{KL}(q_{\phi}(\mathbf{h}|\mathbf{z}) || p(\mathbf{h})) = \mathbf{\mu}^H\mathbf{\mu} + \left|\left| \mathbf{\sigma} - \mathbf{1} - \frac{1}{2}\log(\mathbf{\sigma}^2 - |\mathbf{\delta}|^2) \right|\right|_1
                $$

            \vspace{1em}
            \noindent \textbf{The Reparametrization Trick}
            \vspace{0.5em}

                Even though we have the KL-Divergence, we still need to handle the reparameterization trick to be able to sample from the posterior distribution while also allowing gradients to flow. We begin by decomposing the complex latent vector $\mathbf{h}$ into its real and imaginary components, $\mathbf{x} \in \mathbb{R}^N$ and $\mathbf{y} \in \mathbb{R}^N$, such that $\mathbf{h} = \mathbf{x} + i\mathbf{y}$. Given the assumption of diagonal covariance and pseudo-covariance matrices defining the posterior distribution, all complex variables are mutually independent. Consequently, each individual complex variable $h_j \in \mathbf{h}$ can be modeled as a bivariate real Gaussian distribution, defined as $\mathbf{h}_j = [x, y]^T$, where $\mathbf{h}_j \sim \mathcal{N}(\mathbf{\mu}_{h_j}, \mathbf{\Sigma}_{h_j})$.

                To apply the reparameterization trick, we require a transformation matrix $\mathbf{L}$ such that sampling from a standard bivariate normal distribution, $\mathbf{\epsilon} = [\epsilon_r, \epsilon_i]^T \sim \mathcal{N}(\mathbf{0}, \mathbf{I})$, allows us to express the latent variable as $\mathbf{h}_j = \mathbf{\mu}_{h_j} + \mathbf{L}\mathbf{\epsilon}$. By linear transformation properties, the covariance of this transformed variable is given by $\text{Cov}(\mathbf{L}\mathbf{\epsilon}) = \mathbf{L} \text{Cov}(\mathbf{\epsilon}) \mathbf{L}^T$. Since $\text{Cov}(\mathbf{\epsilon}) = \mathbf{I}$, this simplifies to $\mathbf{L}\mathbf{I}\mathbf{L}^T = \mathbf{L}\mathbf{L}^T$. Thus, we must solve for $\mathbf{L}$ such that $\mathbf{L}\mathbf{L}^T = \mathbf{\Sigma}_{h_j}$.
                
                This is solved using the Cholesky decomposition, which is the matrix equivalent of finding a square root. Because $\mathbf{L}$ must be a lower-triangular matrix, we set it up as 
                
                $$\mathbf{L} = \begin{bmatrix} l_{11} & 0 \\\\ l_{21} & l_{22} \end{bmatrix}$$
                
                Multiplying $\mathbf{L}$ by its transpose $\mathbf{L}^T$ gives us:
                
                $$\mathbf{L}\mathbf{L}^T = \begin{bmatrix} l_{11} & 0 \\\\ l_{21} & l_{22} \end{bmatrix} \begin{bmatrix} l_{11} & l_{21} \\\\ 0 & l_{22} \end{bmatrix} = \begin{bmatrix} l_{11}^2 & l_{11}l_{21} \\\\ l_{11}l_{21} & l_{21}^2 + l_{22}^2 \end{bmatrix}$$

                Equating this to our previously derived target covariance matrix $\mathbf{\Sigma}_{h_j}$ yields the following system:
                
                $$
                \begin{bmatrix}
                    l_{11}^2 & l_{11}l_{21} \\\\
                    l_{11}l_{21} & l_{21}^2 + l_{22}^2 
                \end{bmatrix} = 
                \begin{bmatrix}
                    \frac{\sigma + \delta_r}{2} & \frac{\delta_i}{2} \\\\
                    \frac{\delta_i}{2} & \frac{\sigma - \delta_r}{2} 
                \end{bmatrix}
                $$
                
                Now we simply solve this system of equations to find the values inside $\mathbf{L}$:
                \begin{itemize}
                    \item $l_{11}^2 = \frac{\sigma + \delta_r}{2} \implies l_{11} = \sqrt{\frac{\sigma + \delta_r}{2}}$

                    \item $l_{11}l_{21} = \frac{\delta_i}{2} \implies l_{21} = \frac{\delta_i / 2}{l_{11}} = \frac{\delta_i}{2\sqrt{\frac{\sigma + \delta_r}{2}}} = \frac{\delta_i}{\sqrt{2(\sigma + \delta_r)}}$

                    \item $l_{21}^2 + l_{22}^2 = \frac{\sigma - \delta_r}{2} \implies l_{22} = \sqrt{\frac{\sigma - \delta_r}{2} - l_{21}^2} = \sqrt{\frac{\sigma - \delta_r}{2} - \frac{\delta_i^2}{2(\sigma + \delta_r)}} = \sqrt{\frac{\sigma^2 - \delta_r^2 - \delta_i^2}{2(\sigma + \delta_r)}} = \sqrt{\frac{\sigma^2 - |\delta|^2}{2(\sigma + \delta_r)}}$
                \end{itemize}
                
                Transforming the standard normal noise vector by $\mathbf{L}$ yields the target covariance structure:
                
                $$
                \begin{bmatrix}
                    x_{shifted} \\\\
                    y_{shifted} 
                \end{bmatrix} = \mathbf{L} 
                \begin{bmatrix} 
                    \epsilon_r \\\\ 
                    \epsilon_i 
                \end{bmatrix} = 
                \begin{bmatrix} 
                    l_{11} & 0 \\\\ 
                    l_{21} & l_{22} 
                \end{bmatrix} 
                \begin{bmatrix} 
                    \epsilon_r \\\\ 
                    \epsilon_i 
                \end{bmatrix}
                $$

                Because $\mathbf{L}$ is lower-triangular, the real component $x_{\text{shifted}}$ depends solely on $\epsilon_r$. Conversely, the imaginary component $y_{\text{shifted}}$ depends on both $\epsilon_r$ and $\epsilon_i$, which intrinsically establishes the required covariance and correlation between the real and imaginary dimensions. By adding the mean vector $\mathbf{\mu}_{h_j} = [\mu_r, \mu_i]^T$, we obtain the final, explicit coordinates for the sampled complex variable: $x = \mu_r + l_{11}\epsilon_r$ and $y = \mu_i + l_{21}\epsilon_r + l_{22}\epsilon_i$.
                
                By factoring $\epsilon_r$ and $\epsilon_i$ into the complex plane ($x + iy$), we recover the complex multipliers $k_r$ and $k_i$ established by Nakashika et al.:

                \begin{itemize}
                    \item $k_r = l_{11} + i l_{21} = \sqrt{\frac{\sigma + \delta_r}{2}} + i \frac{\delta_i}{\sqrt{2(\sigma + \delta_r)}} = \frac{\sigma + \delta_r + i\delta_i}{\sqrt{2(\sigma + \delta_r)}}$

                    \item $k_i = i l_{22} = i\sqrt{\frac{\sigma^2 - \vert{}\delta\vert{}^2}{2(\sigma + \delta_r)}}$
                \end{itemize}
                
                Generalizing this scalar derivation to the entire latent space, the final reparameterized complex latent vector $\mathbf{\tilde{h}}$ is expressed as:
                
                $$
                \mathbf{\tilde{h}} = \mathbf{\mu} + \mathbf{k}_r \mathbf{\epsilon}_r + \mathbf{k}_i \mathbf{\epsilon}_i
                $$
                
            \vspace{1em}
            \noindent \textbf{Reconstruction Loss}
            \vspace{0.5em}

                Following the projection through the complex latent space, the decoder optimizes the maximum likelihood estimation of the data $\mathbf{z}$ given the latent variables $\mathbf{h}$. Assuming the decoder output follows a circularly symmetric complex normal distribution $\mathcal{N}_c(\mathbf{\mu}, \mathbf{I}, \mathbf{O})$, the reconstruction objective reduces to the MSE, directly paralleling the standard real-valued VAE. The probability density function for a circular complex Gaussian of a $D$-dimensional vector $\mathbf{z}$ is defined as:
                
                $$
                p(\mathbf{z}) = \frac{1}{\pi^D \vert{}\mathbf{\Gamma}\vert{}} \exp\left( -(\mathbf{z} - \mathbf{\mu})^H \mathbf{\Gamma}^{-1} (\mathbf{z} - \mathbf{\mu}) \right)
                $$
                
                Because we defined the covariance $\mathbf{\Gamma}$ as the identity matrix $\mathbf{I}$, the equation simplifies to:
                
                $$
                p_\theta(\mathbf{z}\vert{}\mathbf{h}) = \frac{1}{\pi^D} \exp\left( -(\mathbf{z} - \mathbf{\mu})^H (\mathbf{z} - \mathbf{\mu}) \right)
                $$
                
                Taking the natural logarithm yields the log-likelihood:
                
                $$
                \log p_\theta(\mathbf{z}\vert{}\mathbf{h}) = \log\left( \frac{1}{\pi^D} \right) - (\mathbf{z} - \mathbf{\mu})^H (\mathbf{z} - \mathbf{\mu})
                $$
                $$
                \log p_\theta(\mathbf{z}\vert{}\mathbf{h}) = -D\log(\pi) - (\mathbf{z} - \mathbf{\mu})^H (\mathbf{z} - \mathbf{\mu})
                $$
                
                During gradient-based optimization, constant terms independent of the network parameters can be disregarded. Since the term $-D\log(\pi)$ produces a zero gradient with respect to the network weights, it is omitted from the loss function. Furthermore, the inner product of a complex vector with its conjugate transpose corresponds exactly to the squared $L_2$ norm. Thus, the proportional log-likelihood is expressed as:
                
                $$
                \log p_{\theta}(\mathbf{z} \vert{} \mathbf{h}) \propto {-||\mathbf{z} - \mathbf{\mu}||_2^2}
                $$

                Evaluating the expectation over the approximate posterior using a single-sample Monte Carlo estimation, analogously to the standard VAE once again, we get:
                
                $$
                \mathbb{E}_{q_\phi(\mathbf{h} \vert{} \mathbf{z})}[\log p_\theta(\mathbf{z} \mid \mathbf{h})] \approx -||\mathbf{z} - \mathbf{\mu}||_2^2
                $$
                
                This confirms that maximizing the complex log-likelihood is mathematically equivalent to minimizing the squared Euclidean distance between the target complex spectrogram and the network's reconstruction.

            \vspace{1em}
            \noindent \textbf{Architecture and Implementation}
            \vspace{0.5em}

                With the mathematical framework fully established, translating these complex-valued constraints into a functional neural network required specialized tooling. For our implementation, we utilized the complexPyTorch library\footnote{https://github.com/wavefrontshaping/complexPyTorch} \cite{complexpytorch, trabelsi2018deep} and built directly upon the architectural foundations developed by Camara et al.\footnote{https://github.com/MateoCamara/complex-vae} \cite{camara2022phase}. 
                
\begin{figure}[h]
                    \centering
                    \includegraphics[width=0.8\textwidth]{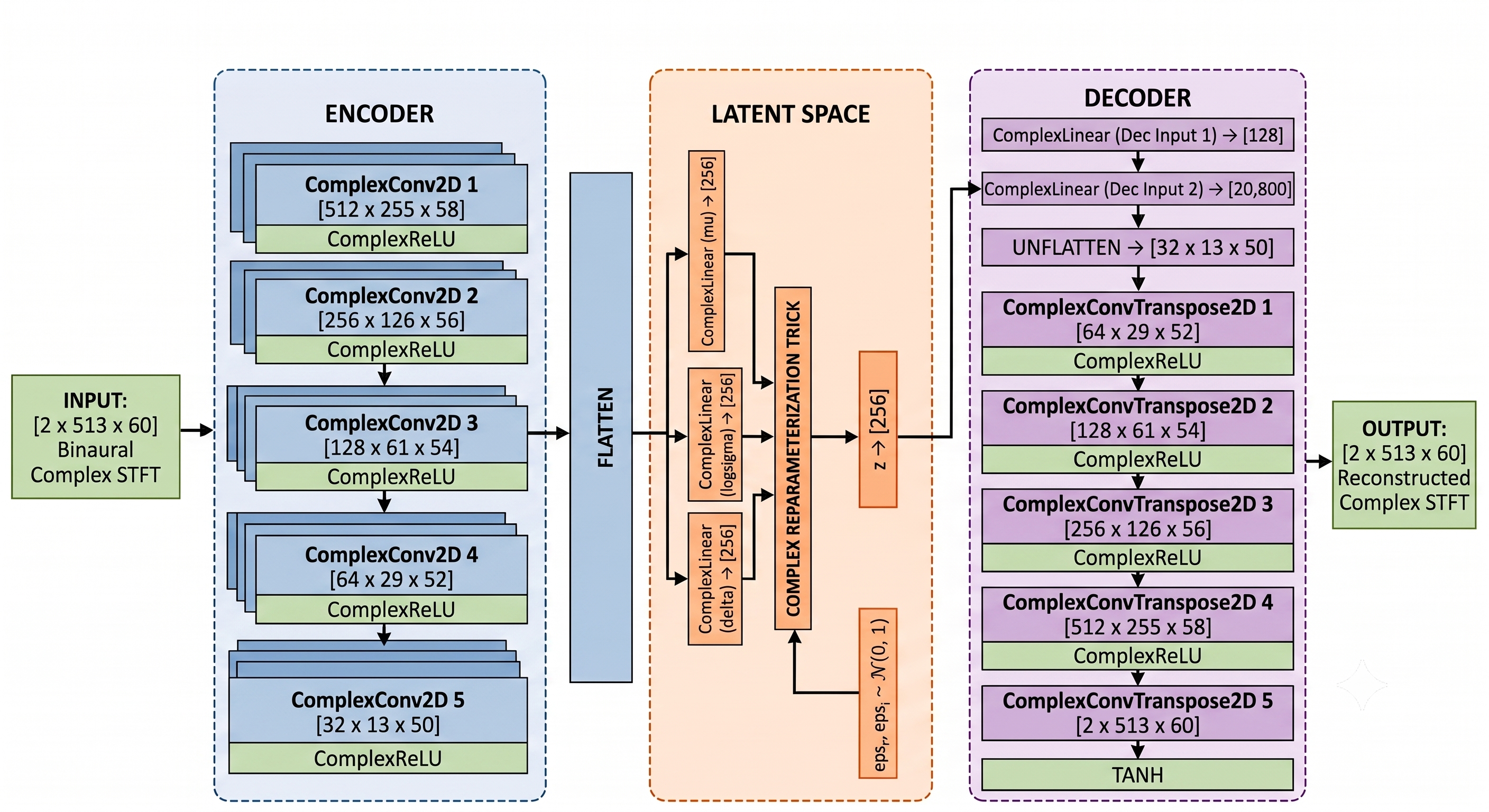}
                    \caption{Architecture overview of the CVAE network, taking binaural complex STFT as input. The encoder comprises five complex 2D convolutional layers, each paired with batch normalization (omitted for visual clarity) and complex ReLU activations. The features then feed into three parallel complex linear branches with 256 dimensions, predicting the means ($\mathbf{\mu}$), log-variances ($\log \mathbf{\sigma}$), and pseudo-variances ($\mathbf{\delta}$). Following the complex reparameterization trick using auxiliary real and imaginary standard normal noise distributions $\mathbf{\epsilon}_r$ and $\mathbf{\epsilon}_i$, the sampled latent vector is passed to the decoder. The decoder consists of two complex linear layers followed by five complex 2D transposed convolutions with batch normalization and complex ReLU activations. The final layer utilizes a real-valued Tanh activation to map the reconstructed real and imaginary coefficients of the output strictly back to the $[-1, 1]$ range.}
                    \label{fig:cvae}
                \end{figure}

                An overview of the CVAE network is illustrated in Figure \ref{fig:cvae}. While the architecture is fully generalizable and scalable for alternative configurations, this complex-valued model is significantly more sensitive to hyperparameter choices and substantially harder to train than its real-valued counterparts. Researchers modifying this pipeline must exercise caution, as improper tuning easily induces catastrophic latent space collapse.
                
    \subsection{Training}
        
        The framework supports both single-model training and parallelized hyperparameter grid searches. The dataset, consisting of roughly six hours of spatial audio trajectories, is partitioned using a 90/10 training-to-test split; the resulting training set is then further divided into a 90/10 training-to-validation split. The standard training loop is configured for 100 epochs. For the Mel and 4-channel STFT VAEs, a batch size of 256 is utilized, consuming approximately 12 GB and 14 GB of VRAM, respectively, on an NVIDIA RTX 4090 GPU. Conversely, the memory-intensive complex-valued VAE (CVAE) requires the batch size to be drastically reduced to 32, which still occupies approximately 24 GB of VRAM.
        
        Training times vary significantly across architectures: the Mel VAE trains in roughly 5 hours, the 4-channel STFT VAE in 10 hours, and the CVAE, assuming early stopping triggers at approximately 25 epochs, in 40 hours, scaling up to 160 hours if run to the maximum 100-epoch limit. Optimization is driven by the Adam optimizer, configured with a learning rate of $1 \times 10^{-3}$ and a weight decay of $1 \times 10^{-5}$. To stabilize the initial training phase -- particularly crucial for the CVAE -- a warmup period of one epoch is enforced. Training is also monitored via an early stopping mechanism with a patience tolerance of 0.02.
        
        A critical component of our training regime is the scheduling of the KL-divergence loss $\beta$ coefficient. The pipeline utilizes a maximum $\beta$ weight of 0.8, modulated through predefined cycles. We found that a linearly increasing start, followed by cyclic oscillatory adjustments, optimally balances the KL-divergence and reconstruction losses. This slow start and cyclic scheduling prevents latent space collapse while still permitting the reconstruction error to decrease steadily over time. Now due to this cyclical nature of the $\beta$ coefficient we impose, periodic oscillations in both the reconstruction and KL-divergence losses are expected; training should persist as long as the macro-level trend of the loss curve remains descending, despite these induced localized fluctuations.
        
        Architectural parameters are strictly enforced to maintain a symmetric encoder-decoder structure. The system allows for dynamic configuration of weight initialization (supporting He, Xavier, or standard PyTorch defaults), the size of the latent space (defined in powers of two), and all convolutional variables, including the number of filters, vertical and horizontal kernel sizes, strides, and padding.
        
        When executing a grid search for hyperparameter tuning, multiple training sessions run concurrently. This demands careful allocation of GPU memory and CPU workers (defaulting to 24 on our 36 core Intel Core i9-10980XE processor) to prevent processing bottlenecks, as well as a judicious selection of the search space size to avoid computationally prohibitive training times. Finally, due to its unique optimization landscape and extreme memory demands, the CVAE required specialized manual tuning and was excluded from the automated grid search.
        
        The trained models for the Mel VAE\footnote{https://huggingface.co/luizerko/binauralvae\_mel}, the 4-channel STFT VAE\footnote{https://huggingface.co/luizerko/binauralvae\_stft\_4ch}, and the CVAE\footnote{https://huggingface.co/luizerko/binauralvae\_mel} are publicly accessible on Hugging Face.

    \subsection{Inference and Dreaming}

        Following training, the system provides an inference module to evaluate the models on unseen navigational trajectories. During inference, the pipeline performs a forward pass through the trained VAE to predict the reconstructed data and subsequently converts these representations back into full spatialized audio waveforms, adhering to the original modality dimensions (for instance, 2048 STFT samples for Mel, 128 Mel bands, and a 147 hop length at 44.1 kHz). For the Mel spectrogram modality, the inherent loss of phase information during preprocessing requires a phase estimation step to synthesize the final waveform. To accommodate this, our pipeline offers two distinct vocoding approaches: the classical Griffin-Lim algorithm \cite{griffinlim} and a modern neural method, BigVGAN-v2\footnote{https://github.com/NVIDIA/BigVGAN} \cite{lee2023bigvganuniversalneuralvocoder}.

        Although the Mel VAE yields high-fidelity spectrogram reconstructions (detailed further in Section \ref{results}), synthesizing the audio without any phase information presents significant challenges. Griffin-Lim reliably produces recognizable audio, although with its well-documented metallic artifacts. For BigVGAN-v2, selected because its pre-trained configuration (44.1 kHz, 128 Mel bands, and a hop length of 256)\footnote{Available on Hugging Face via https://huggingface.co/nvidia/bigvgan\_v2\_44khz\_128band\_256x} most closely align with our acoustic specifications compared to other state-of-the-art vocoders, the resulting audios are typically less recognizable. Without the capacity to fine-tune BigVGAN-v2 to our specific configurations and dataset within our time constraints, the neural vocoder produced audio with excessive noise artifacts, ultimately underperforming the classical approach. Consequently, Griffin-Lim remains the default Mel reconstruction method. In contrast, the 4-channel STFT and complex-valued VAEs natively preserve both magnitude and phase information. For these architectures, the pipeline bypasses phase estimation entirely, recovering the spatial audio waveform directly (and accurately) via the Inverse Short-Time Fourier Transform (ISTFT). Readers are highly encouraged to visit the project repository to listen to and subjectively evaluate audio reconstructions for each method.

        Beyond direct reconstruction, the dreaming module qualitatively evaluates the generative and interpolative capabilities of the learned latent space. By supplying a reference dataset seed -- utilized strictly to extract the correct input tensor dimensions and potentially Mel reference power -- the system synthesizes entirely novel acoustic sequences. It achieves this by calculating a linear interpolation trajectory between a specified number of randomly sampled latent keyframes (requiring a minimum of two). Configured by default to generate 32 interpolated frames between keyframes, this module illustrates how the VAE traverses and transitions between distinct spatial audio states. This generative exercise provides critical insight into the continuity and topological organization of the learned representations, helping to confirm their viability as foundational state descriptors for future audio-based world models. Audio samples of these generated latent trajectories are also available in our repository for auditory review.

\section{Results} \label{results}

% RESULTS:
% MEL: REC -> 19.7783 | KL-DIV -> 7.4318
% STFT-4CH: REC -> 5081.7051 | KL-DIV -> 18.6738
% STFT-COMPLEX: REC -> 625.8284 | KL-DIV -> 4.7277

    Having detailed the proposed pipeline, we now evaluate the objective and subjective performance of the models. The empirical results align closely with our theoretical expectations. All reported metrics are derived from a holdout test set to ensure an unbiased evaluation. Model performance is primarily quantified using two metrics: reconstruction loss, Mean Squared Error (MSE), which indicates the fidelity of the reconstructed output, and Kullback-Leibler (KL) divergence, which measures the difference between the estimated posterior and the prior, serving as an indicator of latent space regularization.

    \subsection{Mel Spectrogram VAE}

\begin{figure}[h]
            \centering
            \includegraphics[width=0.8\textwidth]{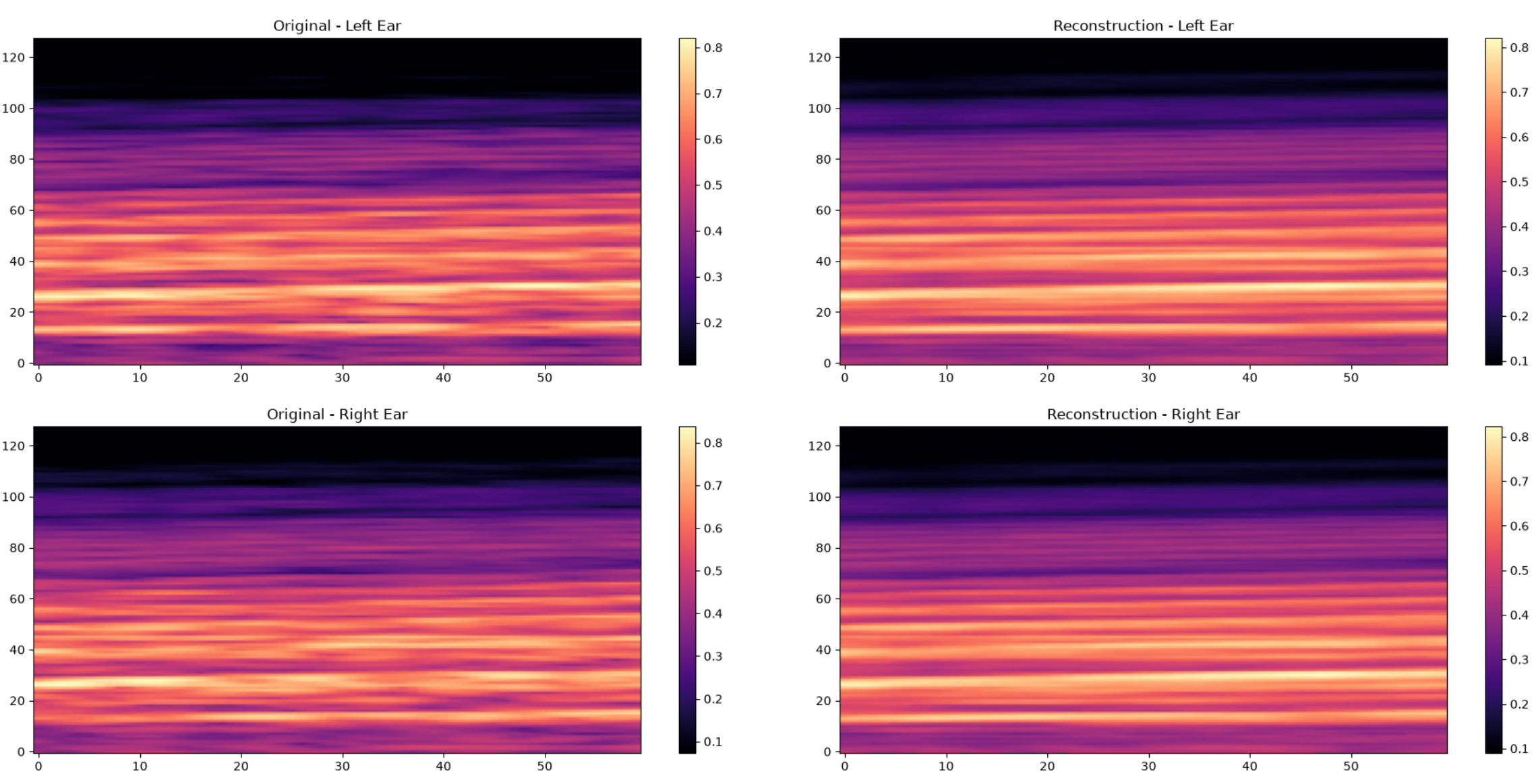}
            \caption{Comparison of the ground truth versus reconstructed Mel spectrogram output from the Mel VAE.}
            \label{fig:mel_reconstruction}
        \end{figure}

        The Mel Spectrogram VAE achieved a reconstruction loss of 19.78 and a KL-divergence of 7.43. These metrics indicate a high-fidelity reconstruction paired with a well-regulated latent space. Because this architecture processes the mel spectrogram as a 2D image, it successfully reproduces the macroscopic structural features of the input. However, the model exhibits the characteristic variance reduction common to Variational Autoencoders (VAEs), resulting in a loss of fine-grained spectral detail. In the audio domain, the synthesized output suffers from the metallic artifacts inherently introduced by phase estimation algorithms during Short Time Fourier Transform (STFT) inversion. Despite this limitation, the reconstructed audio closely mirrors the input perceptually and successfully preserves elements of the original spatialization. Figure \ref{fig:mel_reconstruction} illustrates the visual fidelity of this magnitude-only approach.

    \subsection{4-Channel STFT VAE}

\begin{figure}[h]
            \centering
            \includegraphics[width=0.45\textwidth]{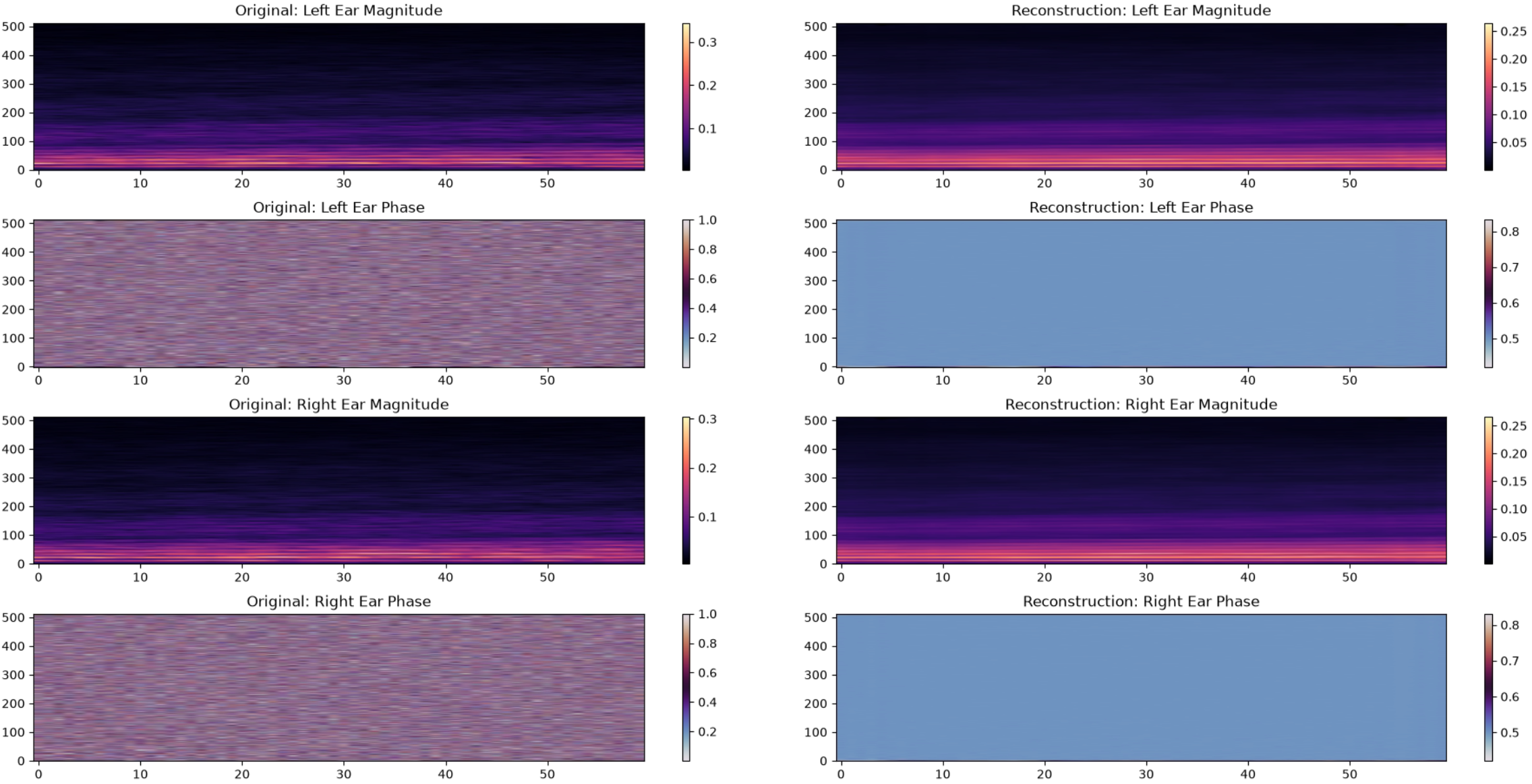}
            \hfill
            \includegraphics[width=0.45\textwidth]{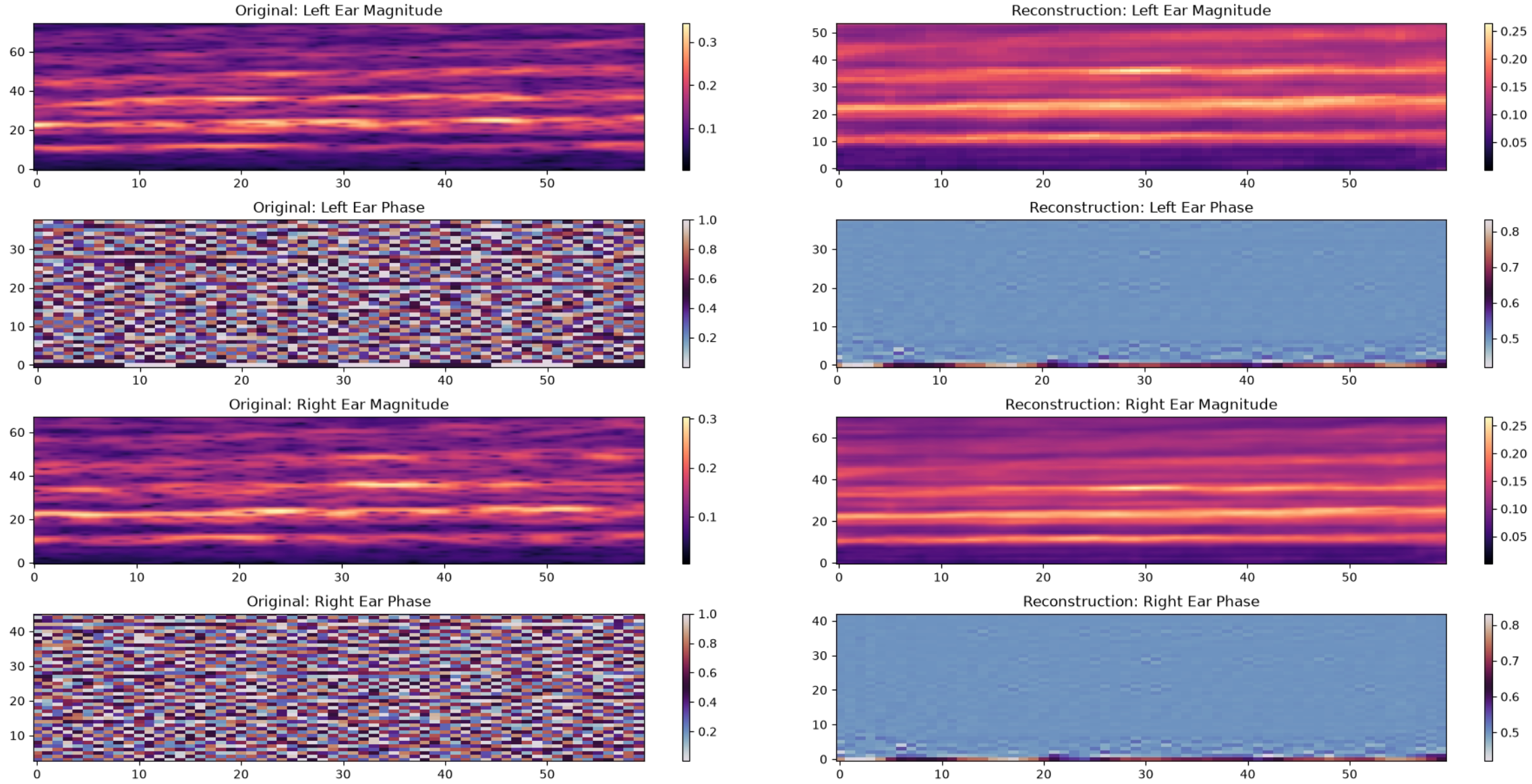}
            \caption{Comparison of the ground truth versus reconstructed spectrogram output from the 4-Channel STFT VAE. To the left, we see the general reconstruction, from a zoom out. To the right, you see the zoomed-in reconstruction on the lower frequencies, where the model performs better, just so we can observe a bit more in detail the accuracy - or lack of it - of the magnitude and phase reconstructions.}
            \label{fig:4ch_reconstruction}
        \end{figure}

        This architecture yielded a significantly higher reconstruction loss of 5081.71, alongside a KL-divergence of 18.67. While the latent space remains stable, the reconstruction fidelity is poor. This model processes both the power-law compressed magnitude and the phase information as spatial image channels. While the model adequately reconstructs the magnitude (although with typical VAE smoothing), representing phase as a spatial image proves highly problematic. To minimize the reconstruction error associated with the highly unstructured phase data, the network regresses to the mean, predicting a near-constant normalized value of approximately 0.5 across the phase image. Consequently, while standard STFT inversion of the ground-truth data achieves perfect reconstruction, the model's failure to predict accurate phase information results in severe signal degradation. The resulting audio retains basic amplitude envelopes (e.g., volume swells) but loses all spatialization and has little perceptual resemblance to the original signal. Figure \ref{fig:4ch_reconstruction} visualizes the results of this approach.

    \subsection{Complex VAE}

\begin{figure}[h]
            \centering
            \includegraphics[width=0.45\textwidth]{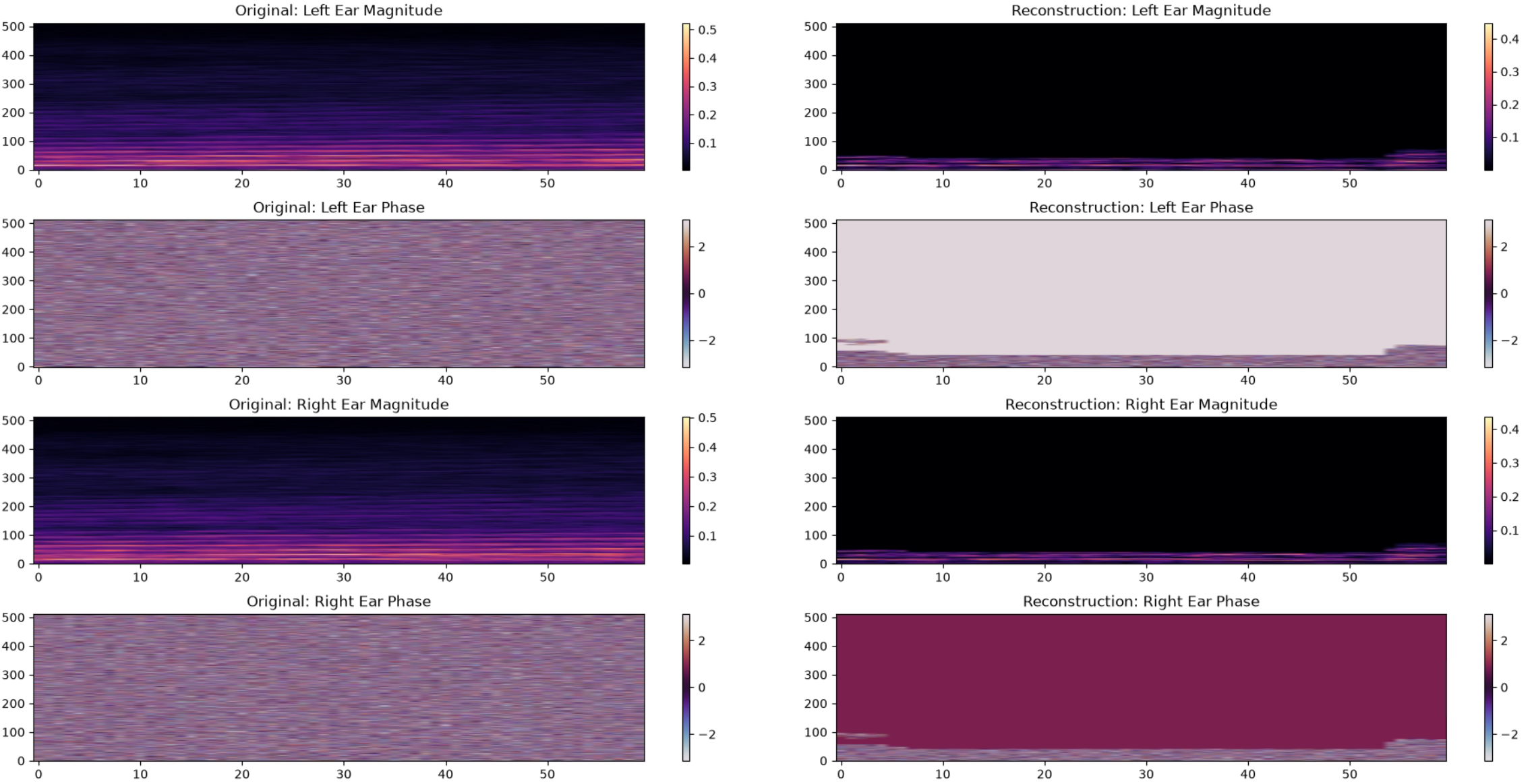}
            \hfill
            \includegraphics[width=0.45\textwidth]{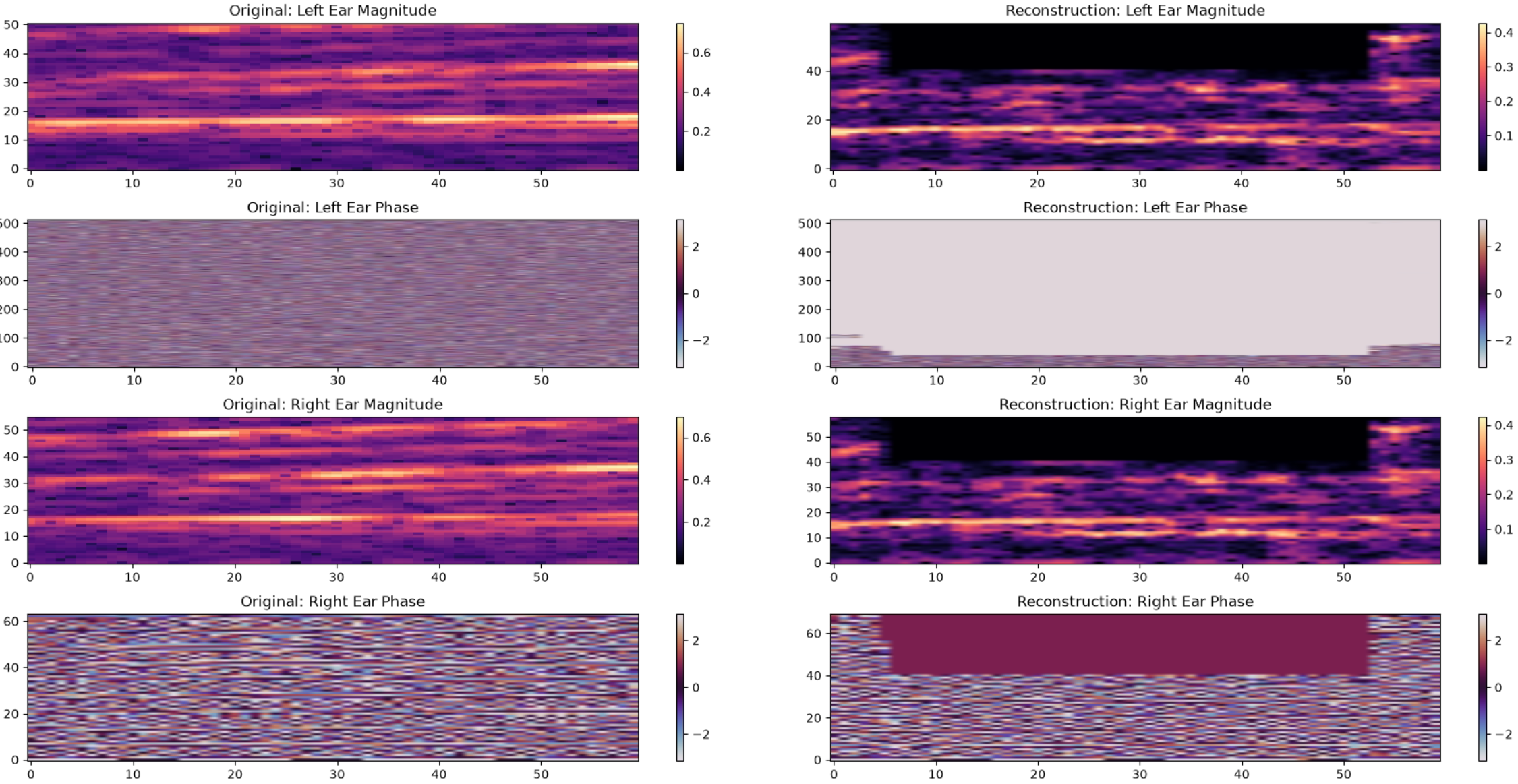}
            \caption{Comparison of the ground truth versus reconstructed spectrogram output from the Complex VAE. To the left, we see the general reconstruction, from a zoom out. To the right, you see the zoomed-in reconstruction on the lower frequencies, where the model performs better, just so we can observe a bit more in detail the accuracy of the frequency and phase reconstructions.}
            \label{fig:cvae_reconstruction}
        \end{figure}

        The Complex VAE achieved a reconstruction loss of 625.83 and a KL-divergence of 4.73, representing a substantial improvement over the 4-Channel STFT VAE, particularly given that it simultaneously reconstructs both magnitude and phase. Treating the input directly as complex numbers, rather than 2D images, fundamentally alters how the VAE's inherent lack of variance manifests. Instead of inducing a uniform visual blur across the spectrogram, the model acts effectively as a low-pass filter. It achieves detailed reconstructions of both magnitude and phase in the lower frequency bands, but struggles to predict high-frequency content. Consequently, while the overall MSE is higher than that of the Mel VAE, the perceptual results are far superior. The accurate low-frequency phase reconstruction allows for robust audio synthesis that preserves the spatialization and fundamental character of the original sound. Perceptually, the output resembles a muffled or filtered version of the original audio, lacking high-frequency texture but maintaining good structural and spatial fidelity. Figure \ref{fig:cvae_reconstruction} visualizes the reconstruction accuracy of this complex-valued approach.
    
\section{Conclusion}

    The challenge of reconstructing spatial audio -- and specifically learning latent representations of spatial audio footprints for navigation and spatial understanding -- remains largely underexplored, particularly within open-source communities and the development of world models. While significant research effort has been dedicated to visual spatial understanding, integrating auditory data represents a natural and necessary progression, as humans inherently rely on acoustic cues for spatial awareness and navigation.

    Our work addresses this gap by presenting a comprehensive pipeline of generalizable Variational Autoencoder (VAE) architectures designed to learn spatial audio latent representations through reconstruction. We contribute to the field by open-sourcing this entire ecosystem, encompassing simulated data acquisition via AudioWorldSim \cite{zerkowski2026audioworldsimrealisticbinauralaudio}, signal processing, model training, latent space sampling, and even pre-trained model weights. By establishing these baselines -- ranging from methods with well-defined limitations to a mathematically grounded complex-valued approach with significant potential -- we aim to provide a solid foundation for future research. We hope the community will leverage and build upon this framework to integrate robust audio-based spatial state representations into future world models and audio-assisted navigation systems.

\section{Limitations and Future Work}

    While this study establishes valuable baselines, several avenues for future research remain. Notably, our most promising architecture, the Complex VAE, requires further investigation and refinement to unlock its full potential. Future efforts should focus on scaling the training dataset, conducting even more extensive architecture and hyperparameter optimization, and investigating the model's training dynamics -- specifically, diagnosing why the training loss tends to converge early, typically between 25 and 30 epochs.

    Furthermore, because this work is primarily motivated by the development of latent audio state representations for future world models, the logical next step is integration into a complete reinforcement learning pipeline, such as an actor-critic framework. Future experiments should evaluate the efficacy of these representations for purely audio-based spatial reasoning tasks, as well as investigate how existing visual world models might benefit from becoming multimodal through the incorporation of our auditory spatial representations.

\section{Ethical Statement and AI Usage Declaration}

    There are no specific ethical concerns, human subject privacy risks, or safety considerations associated with this work. All experiments and datasets are generated entirely within synthetic simulation environments.
    
    Regarding artificial intelligence assistance, Gemini 3.1 chat \cite{GoogleDeepMind2026Gemini31} was utilized during the development of this project to assist with code generation, with direct inclusion of various code snippets created by the AI upon request. Gemini 3.1 chat was also employed during the writing of this technical report. All core conceptual ideas, technical explanations, and an entire draft were human-written, then the AI was utilized for text restructuring and refinement, as well as language revision. Finally, a thorough human revision of every single paragraph took place to get to this final version.

\newpage

\bibliographystyle{alpha}
\bibliography{sample}

\end{document}